\documentclass[11pt,a4paper,twoside]{article}

\usepackage[utf8]{inputenc}
\usepackage[margin=1in,footskip=0.35in]{geometry}
\usepackage{amsmath,amssymb}

\usepackage[super]{cite}
\usepackage{graphicx,caption}
\usepackage{xcolor}
\usepackage[hidelinks]{hyperref}
\usepackage{fancyhdr}

\begin{document}
\noindent
Modern Physics Letters A, Vol.\ \textbf{41}, No.\ 6, 2650011 (2026)\\
\href{https://doi.org/10.1142/S0217732326500112}{DOI: 10.1142/S0217732326500112}
%\vspace{5mm}

% Title, name, address etc.
\begin{center}
\textbf{\LARGE Lorentzian-Euclidean singularity-free\\ solutions to gravitational collapse}\\ \

\vspace{1cm}
Sune Rastad Bahn

\vspace{2mm}
\emph{Gladsaxe Gymnasium, Buddinge Hovedgade 81, DK-2860 S{\o}borg, Denmark}\\
\emph{sunebahn@proton.me}\\ \

Michael Cramer Andersen

\vspace{2mm}
\emph{Christianshavns Gymnasium, Prinsessegade 35, DK-1422 Copenhagen, Denmark}\\
\emph{micran@gmail.com}

\vspace{8mm}
March 4 2026
\end{center}

\vspace{1mm}
\begin{abstract}
This study explores singularity-free solutions to the static, spherical symmetric Einstein equations with the standard Schwarzschild solution as a boundary condition.
Imposing the absence of curvature singularities and requiring differentiability of the time component of the metric leads to a sign change across the horizon, violating the Principle of Equivalence locally. 
We find a solution within the event horizon with a simple ``cosmological constant'' stress-energy tensor. 
Considering the impact of sign change to a compact stellar remnant, modeled by an incompressible perfect fluid obeying the Tolman-Oppenheimer-Volkoff equation, we rediscover the same geometry, indicating both mathematical and physical feasibility of the model.
We also find a new theoretical limit $M/R=3/8$, which is lower than the Buchdahl limit of $M/R=4/9$ for the density of a perfect fluid that will recede behind an event horizon. The equation of state is discussed, and we propose that the final state is described by a Higgs-like free scalar field.\\


%+ 04.70.Bw Classical black holes
%+ 11.30.Cp Lorentz and Poincaré invariance
%+  02.40.Dr Euclidean and projective geometries
%+ 97.60.Jd Neutron stars
%+ 97.60.Lf Black holes
\end{abstract}

%\vspace{-3mm} % To avoid a pagebreak in the itemized list below.

\section{Introduction}
Einstein's theory of general relativity (GR) consists of two independent elements:
\begin{flushleft}
\vbox{
\begin{enumerate}
    \item The Einstein Field Equations (EFE): $G_{\mu\nu}=8\pi T_{\mu\nu}$ connecting the curvature of a manifold with the stress-energy tensor of the matter or energy present.\label{EFE}
    \item The Principle of Equivalence (PE): Along a geodesic (``a free falling observer'') the metric will locally look like the Minkowski spacetime of special relativity.\label{PoE} 
\end{enumerate}
}
\end{flushleft}
It was an early realization that the theory resulted in singularities such as black holes and the Big Bang. One approach to dealing with singularities is wormhole-like solutions.
Another approach to handle the initial big bang singularity, while keeping an origin of time, is to change the signature of the metric. This has been applied both to quantum cosmology by Hartle and Hawking (1983)\cite{HartleHawking1983}, through a rotation to Euclidean time, and to a classical cosmological model by Ellis et al. (1992)\cite{Ellis1992}.   

Hirayama and Holdom (2003)\cite{Hirayama2003}
found black hole solutions with a regular ``Euclidean core'', based on a perfect fluid, in higher derivative theories. 
The term ``Lorentzian-Euclidean black hole'' was introduced by Capozziello et al. (2024)\cite{Capozziello2024} to characterize a signature change across the event horizon, and the geodesics were studied by Bartolo et al. (2025)\cite{Bartolo2025}. 

According to Carballo-Rubio et al. (2020)\cite{Carballo-Rubio2020b}, avoidance of spacetime singularities requires the violation or breakdown of either:
a) the weak energy condition, 
b) the Einstein field equations, 
c) pseudo-Riemannian spacetime geometry, or
d) global hyperbolicity (existence of time-like geodesic complete curves).

In this paper, we consider singularity-free solutions to gravitational collapse while relaxing the constraint (\ref{PoE}) in GR, of ``local Lorentz invariance everywhere'', and find a solution that has a Euclidean core and a regular Lorentzian exterior.
Such solutions obviously violate the ``pseudo'' part of (c). 
Whereas condition (d) of hyperbolicity can be maintained outside the horizon, it becomes inapplicable inside, since there are no causal curves. 
When we consider the matter content of the solution we will insist, not just on (a), but on the stronger \textit{Dominant Energy Condition} corresponding to $\rho\geq |P|$ for a perfect fluid.

% Set the page style to "fancy"...
\pagestyle{fancy}
%... then configure it.
\fancyhead{} % clear all header fields
\renewcommand{\headrulewidth}{0pt}
\fancyhead[RE]{\emph{Lorentzian-Euclidean singularity-free solutions to gravitational collapse}}
\fancyhead[LO]{\emph{S.R.\ Bahn and M.C.\ Andersen}}
% RE (RightEven)
% RO (RightOdd)
% LE (LeftEven)
% LO (LeftOdd)
%

\section{Method of the Study}
We will restrict ourselves to the case of static, spherical symmetric solutions with a ``cosmological constant'' stress energy tensor $8\pi T_{\mu\nu}=-\Lambda g_{\mu\nu}$. We will impose the condition of Minkowski spacetime not as a local condition everywhere for a free-falling observer, but as a boundary condition for $r\to \infty$. We will further restrict ourselves to solutions that have $\Lambda=0$ outside some radius $r_s$.
From Birkhoff's theorem, it is well known that this will lead to the Schwarzschild geometry outside $r_s$. In the interior region, we seek singularity-free solutions with constant $\Lambda$ and a line element that matches the external Schwarzschild line element at the $r_s$ boundary. A smoothness requirement on the solution turns out to imply a signature change.

We proceed to investigate less trivial $T_{\mu\nu}$ cases and find that the solution can be seen as the limiting case of the gravitational collapse of an incompressible perfect fluid, e.g. a neutron star, or even as the solution for a scalar ``Higgs'' field.  

\section{Solving the Einstein Equations}\label{SolvingE}
Consider the general form of a static, spherical symmetric line element (here and elsewhere we use natural units $G=c=1$), with $d\Omega^2=d\theta^2+\sin^2\theta\,d\phi^2$:
\begin{equation}\label{General-line-element}
ds^2=f(r)\,dt^2+h(r)\,dr^2+r^2\,d\Omega^2.
\end{equation}
The temporal and radial components of the Einstein tensor are calculated:
\begin{equation}
    G_{tt}=\frac{f(r)(h(r)-h^2(r)-h'(r)r)}{h^2(r)r^2}  
\end{equation}
and
\begin{equation}
    G_{rr}=\frac{f'(r)r+f(r)-h(r)f(r)}{f(r)r^2}.
\end{equation}
If we now assume a constant stress-energy matrix $T_{\mu\nu}=-\frac{\Lambda}{8\pi} g_{\mu\nu}$ we can solve the Einstein field equation,
\begin{equation}
    G_{tt}=\frac{f(r)(h(r)-h^2(r)-h'(r)r)}{h^2(r)r^2}=8\pi T_{tt}=-\Lambda f(r) 
\end{equation}
resulting in (assuming $f\neq0$), where $c_1$ is an integration constant determined by the boundary conditions:
\begin{equation}
    h(r)=\frac{r}{r-c_1-\frac{1}{3}\Lambda r^3},
\end{equation}
We can now solve for the $rr$ part, 
\begin{equation}
G_{rr}=\frac{f'(r)r+f(r)-h(r)f(r)}{f(r)r^2}=8\pi T_{rr}=-\Lambda h(r) 
\end{equation}
resulting in, where $c_2$ is another integration constant:
\begin{equation}
f(r)=\frac{3c_2(r-c_1-\frac{1}{3}\Lambda r^3)}{r}=\frac{3c_2}{h(r)},
\end{equation}
The resulting line element is thus:
\begin{equation}\label{Schwarzschild-deSitter}
ds^2=3c_2\left(1-\frac{c_1}{r}-\frac{\Lambda}{3} r^2\right)\,dt^2
    +\left(1-\frac{c_1}{r}-\frac{\Lambda}{3} r^2\right)^{-1}\,dr^2+r^2\,d\Omega^2. 
\end{equation}
This is the Schwarzschild-de Sitter metric if $c_2=-1/3$ and $c_1=r_s$. We will, however, not restrict ourselves to this choice of parameters, but be guided by the curvature invariants:
\begin{itemize}
\item Ricci curvature scalar $R=4\Lambda$ (independent of $c_1$ and $c_2$) 
\item Kretschmann curvature scalar $w_1=R_{ijkl}R^{ijkl}=\frac{3c_1^2}{2r^6}$ (indep.\ of $c_2$ and $\Lambda$)
\item determinant $\det g = 3c_2r^4\sin^2\theta$ (indep.\ of $c_1$ and $\Lambda$).
\end{itemize}
Notice that $w_1$ diverges at $r=0$ for $c_1 \neq 0$. The term $c^2=|3c_2|$ can be thought of as a ``speed of light'' scaling factor between the radial and temporal coordinates.

\subsection{The Schwarzschild solution} 
Since we are working in natural units we impose the speed of light $c=1$ outside $r_s$, that is, $c_2=-1/3$, where the sign is to ensure the Lorentzian signature. Also, since we are in the flat case, the cosmological constant is $\Lambda=0$. 
Finally, $c_1=r_s$ is chosen to obtain the usual Schwarzschild solution valid for $r>r_s$. We have: 
\begin{equation}\label{Schwarzschild}    
f(r)=-\frac{r-r_s}{r},
\qquad 
h(r)=\frac{r}{r-r_s}.
\end{equation}
Note that with these choices $f(r_s)=0$ and $f'(r_s)=-1/r_s$. Applying the same constants for $r<r_s$ recovers the inner Schwarzschild solution, with its diverging $w_1$. 
However, we are interested in avoiding the central singularity.

\subsection{A singularity-free solution}
\label{SectionSingularity-freeSolution}
The constants within $r_s$ are determined as follows.
Since the Kretschmann curvature scalar diverges at $r=0$ for $c_1\neq 0$, we will require $c_1=0$ for $r<r_s$.
For $r<r_s$, we also no longer assume $\Lambda=0$. Instead, we impose the condition that the metric matches ($f(r_s)=0$) at $r=r_s$ which can be obtained (for $c_1=0$) using $\frac{1}{3}\Lambda r_s^2=1$, that is, $\Lambda=3/r_s^2$. To determine $c_2$, we require $f$ to be differentiable with continuous $f'(r)=3c_2(-2r/r_s^2)$ across the horizon, therefore $f'(r_s)=-1/r_s$; that is $c_2=1/6$. 

\begin{figure}[htb]
\captionsetup{width=14.0cm}
\centerline{\includegraphics[width=14cm]{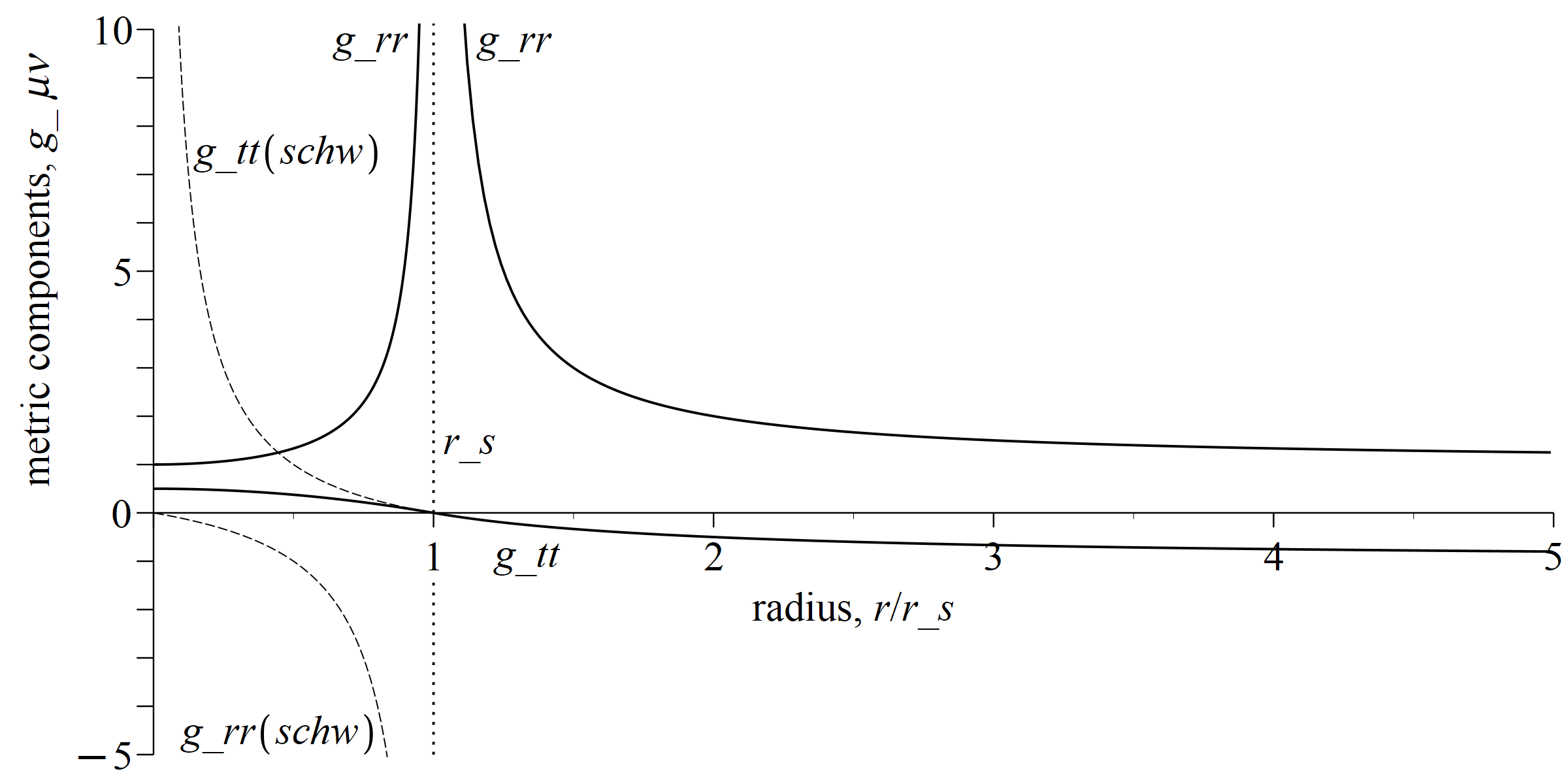}
}
\caption{Components of the metric in equation (\ref{lineelement}) shown with solid curves. The singular inner parts of the Schwarzschild metric are shown with thin dashed curves for comparison.}
\label{Figure-metric}
\end{figure}

The total line element for a singularity-free interior is thus:
\begin{equation}\label{lineelement}  
    ds^2=\begin{cases}
        -\frac{r-r_s}{r}dt^2+\frac{r}{r-r_s}dr^2+r^2d\Omega^2 & \text{for} \quad r_s<r \\
        \frac{1}{2}\left(1-\frac{r^2}{r_s^2}\right)dt^2+\left(1-\frac{r^2}{r_s^2}\right)^{-1}dr^2+r^2d\Omega^2 & \text{for} \quad 0 \leq r< r_s. 
    \end{cases}
\end{equation}
Figure \ref{Figure-metric} plots the metric components compared with the inner Schwarzschild metric.

\vspace{1mm}

The metric connects the internal Euclidean space of constant positive curvature $R=12/r_s^2=3/M^2$ parabola in a smooth way with the external hyperbolic Schwarzschild metric corresponding to a perceived mass of $M=r_s/2$. 

The Euclidean core can be thought of as a de Sitter like micro cosmos by interpreting the $\Lambda$-term in the line element (\ref{Schwarzschild-deSitter}) as a cosmological horizon, $r_H=c/H_\Lambda=\sqrt{3/\Lambda}$.  

The causal structure of the external part of the solution is well established. The metric satisfies hyperbolicity with causal (i.e. null or time-like) geodesics that reach the horizon in finite eigentime. Inside the horizon there are only acausal (spacelike) geodesics. As a consequence, causal geodesics will effectively terminate on the horizon.\cite{Bartolo2025} 

Although the line element~(\ref{lineelement}) is the result of a purely formal derivation, we will see below that this geometry naturally appears from a gravitational collapse if we relax the condition \ref{PoE} of GR.

\section{Final State of Gravitational Collapse }\label{DarkStar}
When studying stellar evolution for stars that undergo gravitational collapse, the usual starting point is, in addition to GR, an equation of state (EoS) for the star, e.g. $P=\omega\rho^\gamma$ with pressure $P$, density $\rho$ and adiabatic index $\gamma$. The associated stress-energy tensor for a relativistic perfect fluid is:
\begin{equation}
    T^{\mu\nu}=(\rho+P)u^\mu u^\nu+Pg^{\mu\nu}.
\end{equation}
Furthermore, if we impose the hydrostatic equilibrium condition, we arrive at the Tolman-Oppenheimer-Volkoff equation (following Section 23.5 in \textit{Gravitation}\cite{MTW1973}):
\begin{equation}
    \frac{dP}{dr}=-\frac{(\rho+P)(m(r)+4\pi r^3P)}{r(r-2m(r))},
\end{equation}
with $m$ being the ``mass'' within the radius $r$,
\begin{equation}
    m(r)=\int_0^r4\pi r^2\rho\ dr.
\end{equation}
\subsection{Incompressible perfect fluid solution}
In the extreme case of $\gamma=\infty$ we have a uniform density $\rho=\rho_0$ that corresponds to the EoS of an incompressible fluid. We can then solve for $P(r)$ for a given radial extension of the star $R$ outside which $\rho$ and $P$ vanishes:
\begin{equation}\label{RadialPressure}
P(r)=\rho_0\frac{\sqrt{1-D\frac{r^2}{R^2}}-\sqrt{1-D}}{3\sqrt{1-D}-\sqrt{1-D\frac{r^2}{R^2}}},   
\end{equation}
with $D=\frac{2m(R)}{R}=\frac{8\pi}{3}\rho_0R^2=\frac{2M}{R}$ twice the ``average mass of the shells''. In the center of the star this reduces to $P(0)=\rho_0\frac{1-\sqrt{1-D}}{3\sqrt{1-D}-1}$ which diverges at the Buchdahl limit of $M/R=4/9$. The physical interpretation of this divergence is that as the star becomes more massive and/or dense, the gravity increases, leading to an increasing central pressure in order to support the star. 
At some point gravity becomes so large that no pressure can withstand it, and the star collapses.  
After the collapse, it is standard to describe the geometry by the Schwarzschild geometry with its associated curvature divergence at the center $r=0$, which is widely regarded as an unphysical singularity.

\subsection{Singularity-free dark compact object solution}
In the spirit of relaxing the condition \ref{PoE} of GR, let us investigate what would happen if we allowed the metric to go positive definite at $r=0$ for some $D$, keeping all other aspects constant.
Upon change of signature, we keep the form of the stress-energy tensor, but the change of $g$ would lead to a change in sign in the coordinates of the $rr$ component, which in the rest frame of the perfect fluid would correspond to a change in sign for the time component $\rho$.
The Euclidean TOV equation hence becomes:
\begin{equation}
    \frac{dP}{dr}=\frac{(\rho-P)(m(r)+4\pi r^3P)}{r(r-2m(r))},
\end{equation}
which puts an upper limit $P\leq \rho$ on the pressure since $dP/dr=0$ when $P=\rho$. This limit corresponds to the Dominant Energy Condition, which could be the physical principle underlying the signature change.
Using the pressure $P(r)$ in equation (\ref{RadialPressure}) for an incompressible fluid, this limit is reached at $r=0$ when $D=3/4$, i.e. $M/R=3/8$, which is roughly $16\%$ smaller than the standard Buchdahl limit of $M/R=4/9$. 

A possible physical interpretation is that as the pressure grows inside a dense stellar core, at some point the pressure component becomes equal to the energy density $P=\rho_0$, at which point the stress energy is strong enough to flip the sign of the geometry to uphold the Dominant Energy Condition, leading to an Euclidean inner core. 

The geometry right outside the Euclidean inner core will look like a Schwarzschild event horizon and will hence absorb all mass, eventually leading to a solution with constant curvature inside some radius and empty space outside.
In other words, we find that the resulting geometry matches exactly the combined solution (\ref{lineelement}) found above with $\Lambda=8\pi\rho_0$.

\subsection{Matching the metric across the event horizon}\label{C3-matching}
A procedure to match the interior and exterior of a metric in terms of curvature invariants, associated with the eigenvalues of the Riemann tensor, was described by Gutiérrez-Piñeres and Quevedo (2019)\cite{Gutierrez-Pineres-Quevedo-2019}. They found that in order to match an exterior vacuum solution, such as the Schwarzschild geometry, the eigenvalues must satisfy the sum rule:
$\sum_{i=1}^6 \lambda_i=0$.

For the perfect fluid, in the Lorentzian case, the sum of the eigenvalues is found to be $\frac{3}{2}(P+\rho)$, which leads to the natural conclusion that pressure and density must vanish at the interface between the exterior vacuum and the interior region.
In the Euclidean case, however, as we have seen above, we effectively have a sign change in $\rho$, leading to the condition $\frac{3}{2}(P-\rho)=0$, or that $P=\rho$. This is exactly the condition we expect to govern the transition to Euclidean signature. Not only does this condition ensure that the Dominant Energy Condition is upheld, it also ensures that the interior eigenvalues match those of the exterior.

The above results have been derived under the assumption of an incompressible ($\gamma=\infty$) perfect fluid. For a more realistic EoS with $\gamma=1$ (perfect gas equation) we have $\omega=P/\rho$ ranging from $0$ for ``cold dust'' to $\omega=1/3$ for fully relativistic particles and, even $\omega=1$ for a free ``Higgs'' scalar field.
Under a gravitational collapse we expect the stress-energy to be dominated by larger and larger $\omega$ and the sign change to happen when $\omega=1$ corresponding to $P=\rho$. 

\section{Relation to Other Dark Compact Object Models}
\label{EoS-Dark-objects}
The conditions $P=\rho$ or $\omega=1$ plus a sign change of $T_{tt}$, e.g., $-\rho\to\rho$, resemble the exotic EoS $P=-\rho$ or $\omega=-1$, if observed from the external Lorentzian part. In other words a free Higgs-like scalar field that exists in an Euclidean region appears from the outside as the same EoS which is characteristic for a vacuum scalar field driving inflation.
In our case, rather than introducing exotic states of matter, we avoid singularities by giving up the Principle of Equivalence locally. 
Gravastar models\cite{MazurMottola2001,Cardoso-Pani-2019} apply such an EoS and connect a Schwarzschild exterior with a de Sitter interior by several layers with different EoS.
With a compactness parameter $\epsilon=1-2M/R=1/4$ and a vanishing Kretschmann scalar, our model is placed outside the regions of gravastars and boson stars, as described in Cardoso and Pani (2019)\cite{Cardoso-Pani-2019}.

\section{Comparison with Astrophysical Observations}
The theoretical OV limit is circa $2.0-3.0~M_\odot$ depending on the EoS and also on the angular momentum, see e.g. Lattimer and Prakash (2007)\cite{LattimerPrakash2007} and
Koehn et al. (2025)\cite{Koehn2025} for a recent discussion.
Our prediction of a tighter limit of $M/R=3/8$ implies a lower collapse threshold than predicted by the Buchdahl limit, potentially reducing the OV limit as well.
Observed neutron star masses peak at $1.2-1.5~M_\odot$ from core-collapse supernovae, with a sharp drop toward $\sim2~M_\odot$ due to fewer massive progenitors.
Higher-mass neutron stars can form via binary accretion, with recycled pulsars piling up just below the OV limit. A few well-constrained mass determinations are found above $2~M_\odot$, with PSR J1748-2021B\cite{Pulsar-Discovery,Pulsar-Thesis} reaching $M\gtrsim 2.5~M_\odot$, belonging to the group of rapidly spinning SupraMassive Neutron Stars.
Above the OV limit there is a mass gap, where only very few detections of low-mass ``black hole'' companions in compact binaries are detected (with large uncertainties) in gravitational wave observations.
Our model is consistent with the current data.
See Tauris and van den Heuvel (2023)\cite{Tauris-Heuvel-2023} for an overview of neutron star observations.

\section{Summary and Outlook}
We have found a curvature singularity-free solution of the Einstein equations that describes a compact object with the Schwarzschild metric as a boundary condition outside the horizon.
Within the horizon, a change of the sign for the time components breaks the Principle of Equivalence.
Physical considerations based on hydrostatic equilibrium and the Dominant Energy Condition leads to the same energy-momentum distribution, for gravitational collapse described by an idealized EoS, further corroborating the feasibility of the found solution. 
Finally the realization of the solution as a Higgs-like scalar field is discussed, and compared with other proposed solutions as well as observations.

Further studies could include a stability check, generalization to include angular momentum and/or electrical charge, and an investigation of time dependence and causality.

This work enjoyed no funding from grants. No competing interests exist.

\vspace{-3mm}
\section*{ORCID:} 
Sune Rastad Bahn ID:
\href{https://orcid.org/0009-0009-8350-0054}{https://orcid.org/0009-0009-8350-0054}\\
Michael Cramer Andersen ID:
\href{https://orcid.org/0000-0002-6868-7246}{https://orcid.org/0000-0002-6868-7246}

\end{document}